\documentclass[journal]{IEEEtran}

\usepackage{cite}
\usepackage{array}
\usepackage{color}
\usepackage{float}
\usepackage[cmex10]{amsmath}
\usepackage{nomencl}						%nomenclature
\usepackage[normalem]{ulem}			%underline
\usepackage{diagbox}						%table head
\usepackage{slashbox}						%table head
\usepackage{colortbl}						%table color
\usepackage{multirow}						%table combine
\usepackage{tabularx}
\usepackage{placeins}
\usepackage{algorithm}
\usepackage[noend]{algpseudocode}
\usepackage{siunitx}
\usepackage{url}
\usepackage[hidelinks]{hyperref}

\usepackage{gensymb}

\ifCLASSINFOpdf
  \usepackage[pdftex]{graphicx}
	\graphicspath{{./figure/}}
  \DeclareGraphicsExtensions{.pdf,.jpeg,.png}
\else
  \usepackage[dvips]{graphicx}
  \graphicspath{{./figure/}}
  \DeclareGraphicsExtensions{.eps}
\fi

\ifCLASSOPTIONcompsoc
  \usepackage[caption=false,font=normalsize,labelfont=sf,textfont=sf]{subfig}
\else
  \usepackage[caption=false,font=footnotesize]{subfig}
\fi

\newcommand{\figref}[1]{{Fig.~\ref{#1}}}

\newcommand{\equref}[1]{{(\ref{#1})}}
\newcommand{\sectionref}[1]{{Section~\ref{#1}}}

\newcommand{\highlight}[1]{\textcolor{black}{#1}}
\newcommand{\revision}[1]{\textcolor{black}{#1}}
\newcommand{\smallrevision}[1]{\textcolor{black}{#1}}

\begin{document}
\clearpage
\onecolumn
\thispagestyle{plain}
\twocolumn[
\begin{@twocolumnfalse}
	\begin{center}
	\vspace{4cm}
	\LARGE
	\textbf{Copyright Statements}
	\vspace{1cm}
	\end{center}
	\Large
	This work has been submitted to the IEEE for possible publication. Copyright may be transferred without notice, after which this version may no longer be accessible.
\end{@twocolumnfalse}]

\clearpage
\bstctlcite{IEEEexample:BSTcontrol}
\setcounter{page}{1}

\title{\revision{Mapping of Dynamics between Mechanical and Electrical Ports in SG-IBR Composite Grids}}
\author{Yitong Li, \IEEEmembership{Member, IEEE}, Yunjie Gu, \IEEEmembership{Senior Member, IEEE}, Timothy C. Green, \IEEEmembership{Fellow, IEEE}}
%\thanks{Yitong Li, XXX and XXX are with the Department of Electrical and Electronic Engineering, Imperial College, London. E-mail: yitong.li15@imperial.ac.uk; yunjie.gu@imperial.ac.uk; t.green@imperial.ac.uk.}
%\thanks{This work was supported by...}

\ifCLASSOPTIONpeerreview
	\maketitle %\IEEEpeerreviewmaketitle
\else
	\maketitle
\fi

% ====================================================
% Abstract
% ====================================================

\begin{abstract}

Power grids are traditionally dominated by synchronous generators (SGs) but are currently undergoing a major transformation through the increasing integration of inverter-based resources (IBRs). \revision{The SG-dominated grid is traditionally analyzed in a \textit{mechanical-centric} view which ignores fast electrical dynamics and focuses on the torque-speed dynamics. By contrast, analysis of the emergent IBR-dominated grid usually takes the \textit{electrical-centric} view which focuses on the voltage-current interaction. In this article, a \textit{port-mapping method} is proposed to fill the gap between these approaches and combine them in a unified model. Specifically, the mechanical dynamics are mapped to the electrical impedance seen at the electrical port; and the electrical dynamics are also mapped to the torque coefficient seen at the mechanical port. The bidirectional mapping gives additional flexibility and insights to analyze the sub-system interactions in whole-system dynamics and guide the tuning of parameter.} Application of the proposed method is illustrated in three cases with increasing scales, namely a single-SG-infinite-bus system, a single-IBR-weak-grid system, and a modified IEEE 14-bus SG-IBR composite system.

\end{abstract}

% ====================================================
% Keywords
% ====================================================


% ====================================================
% Section: Introduction
% ====================================================

\section{Introduction} \label{Section:Introduction}

Power systems have traditionally been dominated by synchronous generators (SGs) \cite{kundur1994power,demarco1998design}. In recent years, due the drive towards renewable energy, power electronics and inverter-based resources (IBRs) have become increasingly common features of the power system \cite{blaabjerg2006overview,gu2019transfverter,shair2019overview}, and thus the power system is being transformed from SG-dominated to SG-IBR composite. This structural change results in new oscillation modes and the emergence of new stability problems \cite{shair2019overview,bialek2020does,markovic2021understanding,curi2017control}. 

\begin{figure}[t!]
\centering
	\subfloat[]{\includegraphics[scale = 0.8]{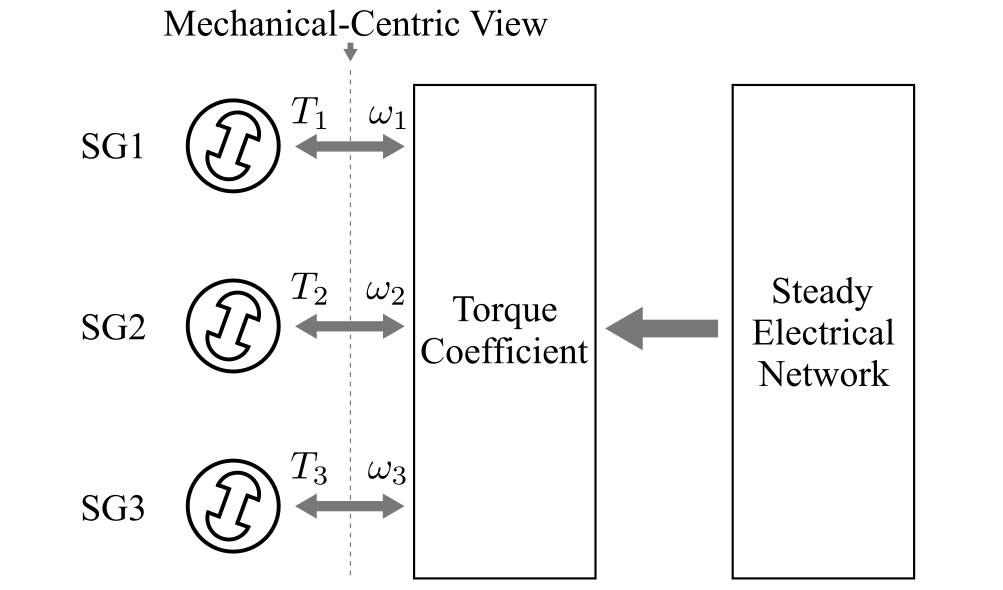}}
	
	\subfloat[]{\includegraphics[scale = 0.8]{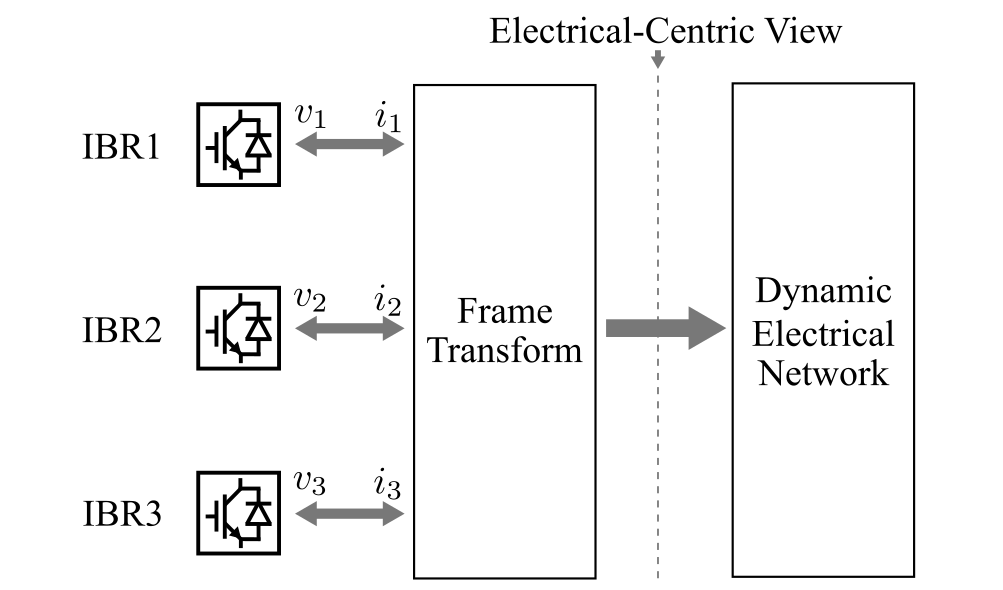}}
	
	\subfloat[]{\includegraphics[scale = 0.8]{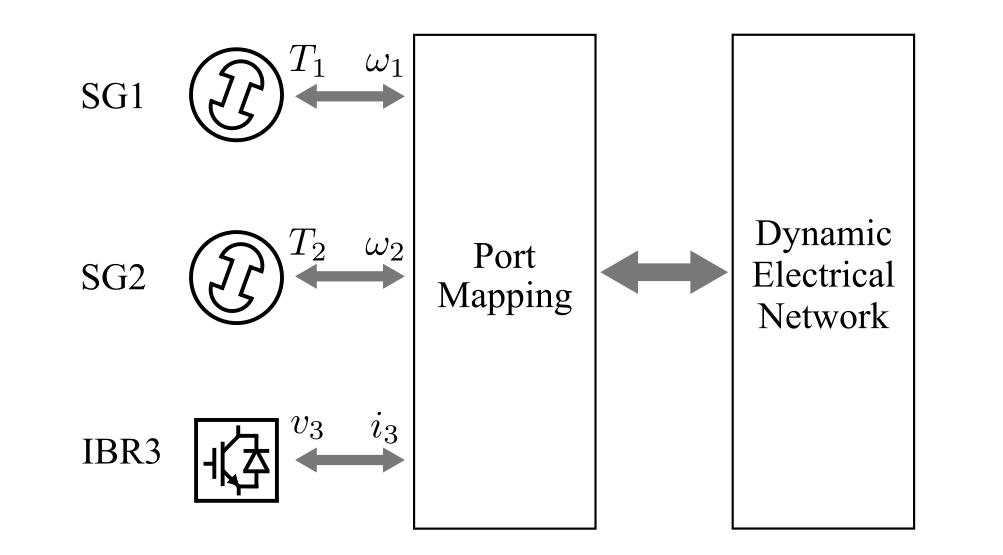}}
\revision{\caption{Power system analysis framework. (a) Mechanical-centric view of SG-dominated systems, where $T$ and $\omega$ are torque and rotor speed respectively. (b) Electrical-centric view of IBR-dominated power systems, where $v$ and $i$ are voltage and current respectively. (c) Port-mapping view of SG-IBR composite power systems.}}
\label{Fig:NetworkModel}
\end{figure}

\revision{Modeling the system dynamics is the first step to addressing the emerging challenges to system stability. One fundamental challenge for an SG-IBR composite grid is that different modeling frameworks have been adopted and evolved for SGs and IBRs. For an SG-based grid, a \textit{mechanical-centric} perspective has been adopted: the speed and rotor angle of synchronous generators are selected as state variables, and the behavior of the grid is considered in terms of the torque-speed ($T$-$\omega$) interaction between the SGs that transfer power via the power flow on transmission lines. For this perspective, the electrical network is transformed to a mechanical representation as torque coefficients, as illustrated in \figref{Fig:NetworkModel} (a) \cite{kundur1994power}. This perspective has been very successful for SG dominated grids, but the fast electrical dynamics of the network are neglected when transformed to torque coefficients, meaning that it is not applicable when fast electromagnetic transients (EMTs) are of concern.} 
 
\revision{In contrast, modelling of an IBR-based grid usually takes an \textit{electrical-centric} perspective in which voltages and currents are the focus of the model. The behavior of the grid is represented by the voltage-current ($v \text{-} i$) interaction between the IBRs and the network, as shown in \figref{Fig:NetworkModel} (b) \cite{middlebrook1989null,liu2003stability,gu2015passivity,harnefors2007modeling,li2021impedance}. This approach preserves all electrical dynamics of IBRs, but it is not straightforward to represent the mechanical dynamics of SGs (torque, speed, and inertia of the rotor) by using the electrical variables (the voltage and current of the stator). Therefore, this method was normally used for super-synchronous (or harmonic) analysis where the synchronization (governed by the rotation of mechanical angles) is not of concern.}

\revision{It is clear that neither approach fully models an SG-IBR composite grid where both EMT and rotational synchronisation make important contributions to whole-system stability of the power system. Therefore, a framework that can unite the mechanical- and electrical-centric views is highly desired. The fundamental difficulty in establishing a unified view is that frame transformations are needed to align the mechanical angles of the various SGs (as well as the phase-locked angles of the various IBRs) in the dynamic electrical network. The frame transformations themselves contain dynamics meaning that SG and IBR in different frames cannot be directly connected \cite{wang2018small}. To address this difficulty, a method called frame dynamics embedding is proposed in \cite{gu2021impedance} where the rotational dynamics in the frame transformation are embedded into electrical models in order that frame transformations do not need separate treatment of dynamics. The frame dynamics embedding method is applied to both SGs and IBRs, yielding impedance element models where the mechanical rotors and the phase-locked loops are represented as equivalent electrical elements \cite{gu2021impedance,li2021impedance,li2020interpreting}. This representation points the way towards a possible unified mechanical-electrical modelling framework capable of predicting the interaction between swing dynamics and inductance dynamics in low-inertia grids.} 

\revision{In this paper, we proposed an inverse approach to the frame dynamics embedding method in \cite{gu2021impedance} by transforming the electrical dynamics back to the mechanical-centric view. This approach yields a dynamic torque coefficient which includes EMT features on top of the classic torque coefficient. The dynamic torque coefficient provides a natural interpretation of IBR-SG interaction from the mechanical viewpoint such that the established concepts and terminology for SG-dominated grids can still be used for grids with high-penetration of IBRs and fast electrical dynamics. The proposed approach, combined with the frame dynamics embedding method, completes the picture of a unified mechanical-electrical view. It is named the \textit{port-mapping method} in this article because it bidirectionally maps dynamics across mechanical and electrical ports, as illustrated in \figref{Fig:NetworkModel} (c). The port-mapping method can also be generalised to other ports, such as the dc-links in IBRs or the exciters in SGs.}

The paper is organized as follows. In \sectionref{Section:PortCouplingMethod}, the concept of port-mapping is introduced and the detailed modelling procedure is illustrated on a SG and a IBR. \sectionref{Section:GeneralPortCouplingMethod} links the port-based models in a SG-IBR composite grid to show the extra flexibility and insights provided by the proposed method in system dynamic analysis. \sectionref{Section:CaseStudy} demonstrated the proposed method in three case studies. \sectionref{Section:Conclusions} concludes the paper.

% =========================
% Section: Single Machine Model
% =========================

\section{The Port-Mapping Method} \label{Section:PortCouplingMethod}

In this section, we introduce the fundamentals of the port-mapping method via the examples of a SG and an IBR. 

\subsection{Mechanical-Electrical Two-Port Network} \label{Section:PortCoupling_TwoPortModel}

\begin{figure}[t!]
\centering
	\subfloat[]{\includegraphics[scale = 1]{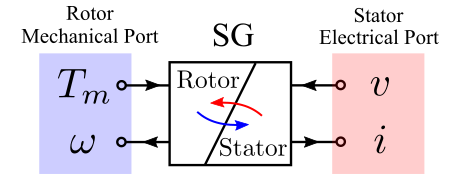}}
	
	\subfloat[]{\includegraphics[scale = 1]{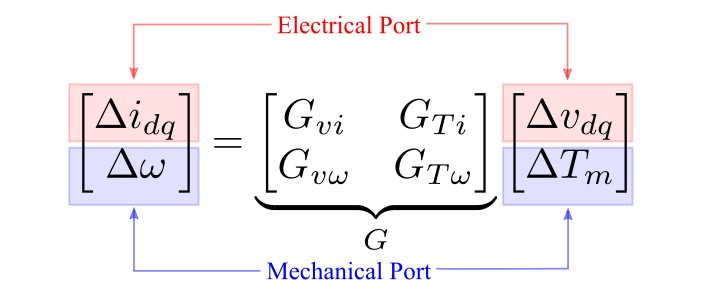}}
\caption{Two-port model of SG. \revision{(a) Port-mapping flow.} (b) Transfer function model.}
\label{Fig:InInteraction}
\end{figure}

Inspired by the two-port network theory in electrical circuits \cite{boylestad2010introductory}, we represent a SG as a mechanical-electrical two-port model as shown in \figref{Fig:InInteraction}. The mechanical port captures the mechanical dynamics between mechanical torque $T_m$ and rotor speed $\omega$ through the rotor shaft, and the electrical port captures the electrical dynamics between terminal voltage $v$ and current $i$ through stator windings. The two ports interact with each other via electrical-mechanical coupling between the rotor and the stator. The mathematical description of the mechanical-electrical two-port network can be formulated as a transfer function matrix $G$ whose diagonal elements represent the self dynamics of a port and the off-diagonal elements represent cross-port coupling, as illustrated in \figref{Fig:InInteraction} (b). The matrix can be readily obtained by linearizing the state equations of a SG as described in Appendix \ref{Appendix:Equ_SG}. 

\revision{The mechanical-electrical two-port model provides a general step for cross-domain modelling but there are difficulties to overcome before these modular models can be used in whole-system analysis. The model of a SG is presented in its local $dq$ reference frame aligned to its rotor, which needs to be aligned  to a global reference frame $DQ$ when the electrical ports of multiple apparatuses are linked in a grid. The frame alignment contains dynamics in itself which is usually represented as extra global synchronising signals, making the linked whole-system model lose modularity. To solve this problem, we transform the electrical port to the global $DQ$ frame and map the frame dynamics into the port matrix so that no extra global synchronisation signals are needed and the whole-system model retains modularity, as described in the following sub-section.}

\subsection{Mechanical Dynamics Mapping to Electrical Ports} \label{Section:PortCoupling_Frame}

\begin{figure*}[t!]
%\centering
\flushleft
	~~~~~~~~~~~
	\subfloat[]{\includegraphics[scale = 1]{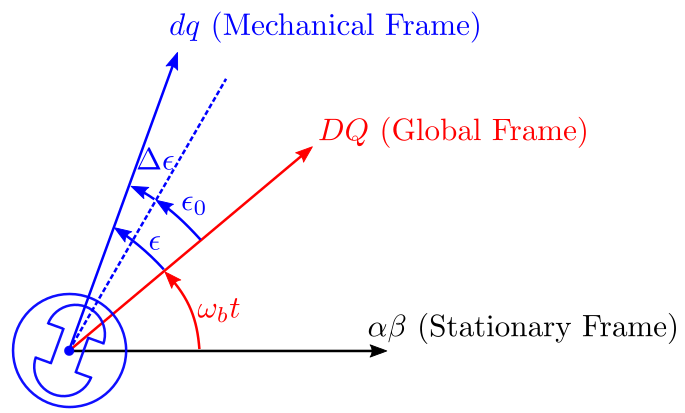}}
	~~~~~~~~~~~~~~~~~~~~~~~~~~
	\subfloat[]{\includegraphics[scale = 1]{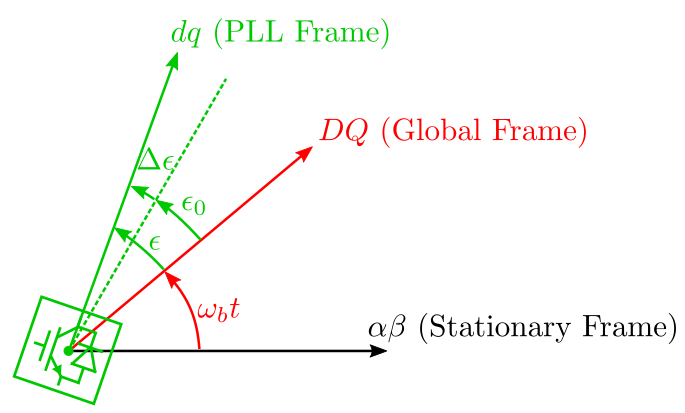}}
	
	~~~~~
	\subfloat[]{\includegraphics[scale=1]{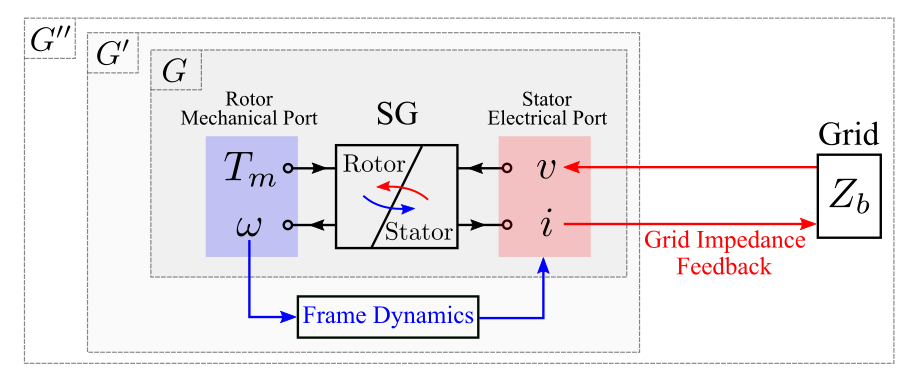}}
	~~~~~~~~~
	\subfloat[]{\includegraphics[scale=1]{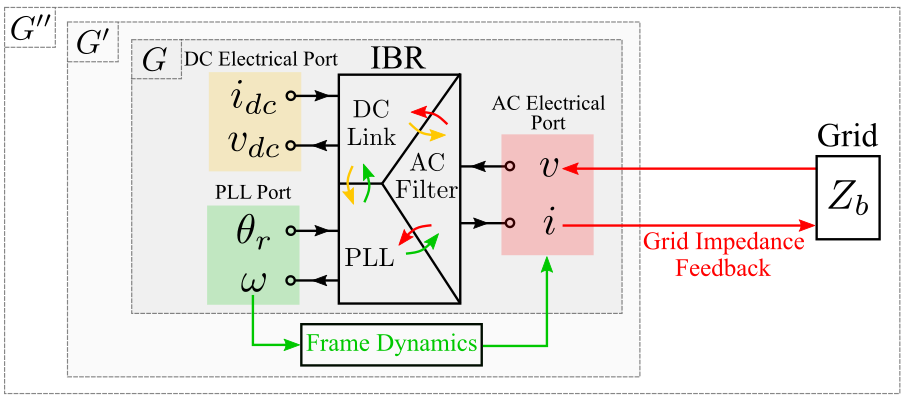}}
	
	\subfloat[]{\includegraphics[scale = 0.67]{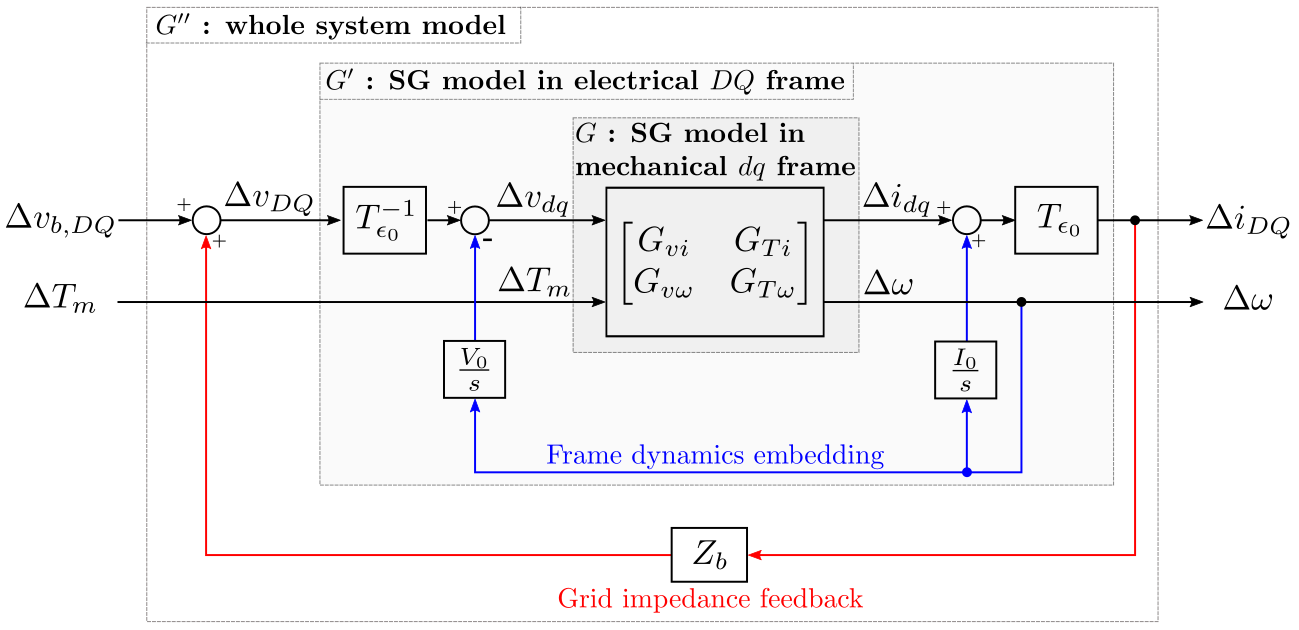}}
	~
	\subfloat[]{\includegraphics[scale = 0.67]{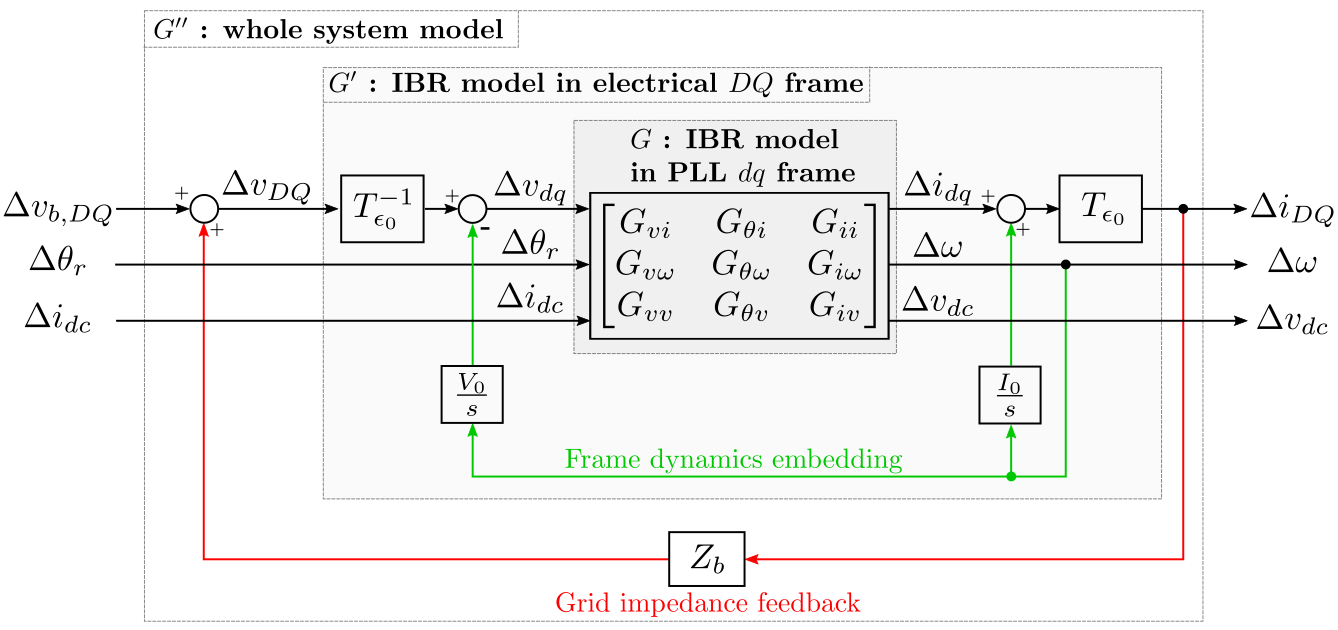}}
\caption{Modeling single-apparatus-infinite-bus system by port-mapping method. (a) Frame illustration of SG. (b) Frame illustration of IBR. \revision{(c) Port-mapping flow of SG.} (d) Port-mapping flow of IBR. (e) Block diagram of SG. (f) Block diagram of IBR.}
\label{Fig:SingleMachineInfiniteBus}
\end{figure*}

Our approach is based on the key observation that the mechanical dynamics of the rotor is identical to the frame dynamics of the electrical port, which offer a bridge for dynamic mapping across mechanical-electrical ports, as illustrated conceptually in the port-mapping flow in \figref{Fig:SingleMachineInfiniteBus} (c) and shown in detail in the block diagram in \figref{Fig:SingleMachineInfiniteBus} (e). We now give the mathematical details. The transformation from mechanical frame $dq$ to global frame $DQ$ for an arbitrary signal $u$ can be represented using the transformation matrix $T_\epsilon$ as
\begin{equation} \label{Equ:FrameDynamics_LargeSignal}
\underbrace{\begin{bmatrix} u_{D} \\ u_{Q} \end{bmatrix}}_{u_{DQ}}
=
\underbrace{\begin{bmatrix} \cos{\epsilon} & -\sin{\epsilon} \\ \sin{\epsilon} & \cos{\epsilon}  \end{bmatrix}}_{T_{\epsilon}}
\underbrace{\begin{bmatrix} u_d \\ u_q \end{bmatrix}}_{u_{dq}}
\end{equation}
where $\epsilon$ is the angle difference between $dq$ and $DQ$ as indicated in \figref{Fig:SingleMachineInfiniteBus} (a). By linearizing this equation, we get
\begin{equation} \label{Equ:FrameDynamics_SmallSignal}
\underbrace{\begin{bmatrix} \Delta u_{D} \\ \Delta u_{Q} \end{bmatrix} }_{\Delta u_{DQ}}
=
T_{\epsilon_0}
\bigg(
\underbrace{\begin{bmatrix} \Delta u_d \\ \Delta u_q \end{bmatrix}}_{\Delta u_{dq}}
+
\underbrace{\begin{bmatrix} -u_{q0} \\ u_{d0} \end{bmatrix}}_{U_0}
\Delta \epsilon
\bigg)
\end{equation}
where the subscript $0$ denotes steady-state operating points and the prefix $\Delta$ denotes perturbation around the operating points. The angle perturbation $\Delta \epsilon$ is the integration of the mechanical speed, that is, $\Delta \epsilon = \Delta \omega / s$ where $s$ is Laplace operator so $1/s$ represents integration. This relationship is indicated by the blue arrows in the block diagram \figref{Fig:SingleMachineInfiniteBus} (e) and yields the transformation from $G$ to $G^\prime$, where the electrical port of $G^\prime$ is represented in the global $DQ$ frame so that it can be readily linked with other apparatuses in the electrical network. The detailed process of the transformation and the element-wise value of $G^{\prime \prime}$ can be found in Appendix \ref{Appendix:Step_Frame}. \revision{The transformed $G^\prime$ has an identical effect to the frame-dynamic embedding method in \cite{gu2021impedance} but preserves the mechanical port on top of \cite{gu2021impedance} so that the bi-directional mapping between the two ports are available. This allow us to map the electrical dynamics back to mechanical ports which provide extra flexibility and insight in stability analysis, to follow.} 

\subsection{Electrical Dynamics Mapping to Mechanical Ports} \label{Section:PortCoupling_SG_Grid}

In the mechanical-electrical two-port model $G^\prime$ the terminal voltage $v_{DQ}$ is taken as an independence input. In reality, however, $v_{DQ}$ is affected by the current $i_{DQ}$ via the equivalent impedance $Z_{b}$ seen at the bus where the apparatus is connected. This yields an extra loop in the block diagram as indicated by the red arrow in \figref{Fig:SingleMachineInfiniteBus} (e). Closing this loop yields the transformation from $G^{\prime}$ to $G^{\prime \prime}$ where the electrical impedance of the grid $Z_{b}$ is mapped into the two-port network. It should be noted that input voltage of $G^{\prime \prime}$ is changed from $v_{DQ}$ to $v_{bDQ}$ which is the equivalent voltage of the grid behind $Z_b$. The detailed process of the transformation and the element-wise value of $G^{\prime \prime}$ can be found in Appendix \ref{Appendix:Step_SG_Grid}.

\subsection{Generalising to IBR}

The port-mapping methodology for a SG discussed in the previous sub-sections can be readily generalised to other apparatuses, such as an IBR, as illustrated in the right column in \figref{Fig:SingleMachineInfiniteBus}. An IBR (grid-following) has a rotating angle in its phase-locked loop (PLL) which is used for synchronisation. The active power is regulated by the balancing of the dc-link voltage. The PLL and the dc-link in a grid following IBR roughly have the same function as the rotor in a SG, for synchronisation and power balancing. Therefore, an IBR is three-port apparatus where the PLL and dc-link port jointly serve as a virtual mechanical port. The port-mapping transformation from $G$ to $G^\prime$ and to $G^{\prime\prime}$ for an IBR can be obtained easily from the block diagrams in \figref{Fig:SingleMachineInfiniteBus} (f).

% =========================
% Section: Network Model
% =========================

\section{Port-Based Analysis of SG-IBR Composite Grids} \label{Section:GeneralPortCouplingMethod}

Having established a step-by-step procedure for dynamic mapping across ports in a SG and a IBR, we now use the derived port-based model to examine the stability of SG-IBR composite grids.

\subsection{Linking SG-IBR Composite Grids}
\begin{figure}[t!]
\centering
\includegraphics[scale=0.9]{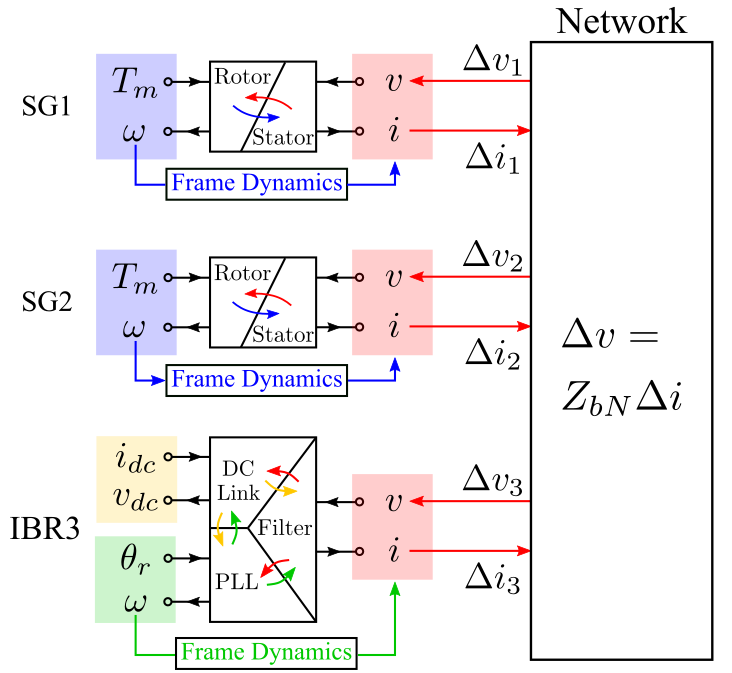}
\caption{Port-mapping flow for a SG-IBR Composite power network.}
\label{Fig:PortBasedNetworkModel}
\end{figure}

\begin{figure*}[t!]
\centering
\includegraphics[scale=0.9]{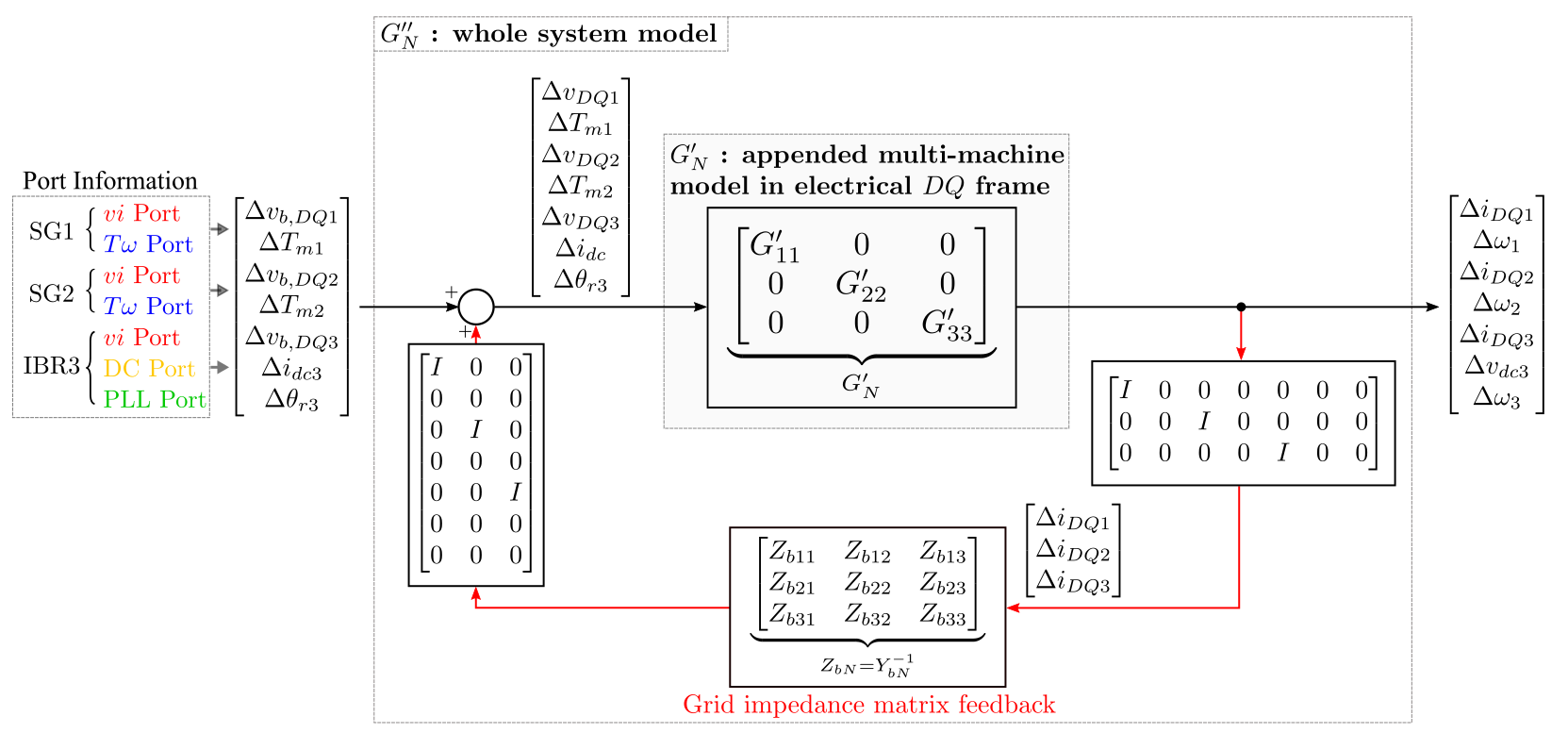}
\caption{Block diagram for the SG-IBR composite power network in \figref{Fig:PortBasedNetworkModel}.}
\label{Fig:BlockDiagram_Network}
\end{figure*}

\begin{figure}[t!]
\centering
\includegraphics[scale=0.9]{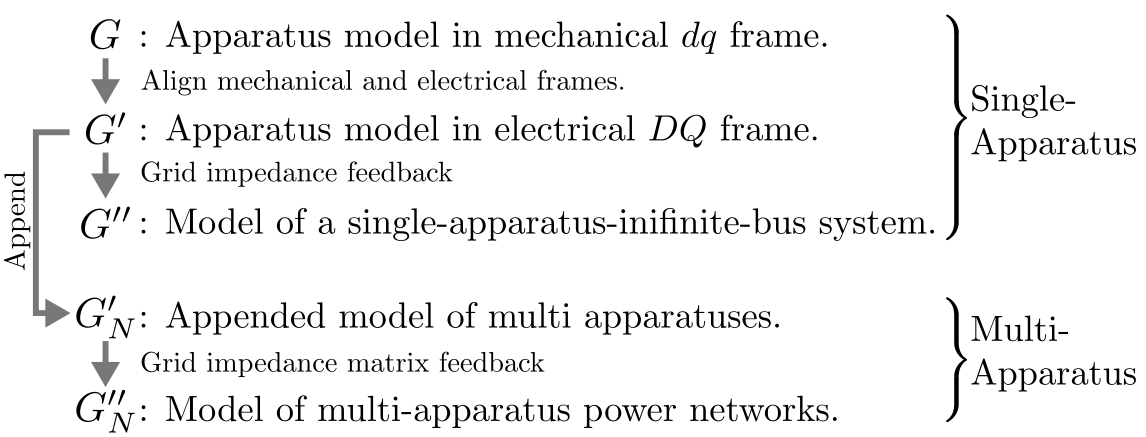}
\caption{Model derivation tree: the procedure to get single-apparatus and multi-apparatus port matrices.}
\label{Fig:ModelDerivationTree}
\end{figure}

We first link the port-based SG and IBR models in a composite network. Taking the network in \figref{Fig:PortBasedNetworkModel} as an example, the corresponding block diagram is shown in \figref{Fig:BlockDiagram_Network}. The port matrix $G$ for each apparatus is transformed to $G^\prime$ (this is done locally without the knowledge on the whole grid), and then appended on $G_{N}^{\prime}$ to represent the collective port matrices of all apparatuses. $G^\prime_N$ interact with the network impedance matrix $Z_{bN}$ to form the whole-system port model $G_{N}^{\prime\prime}$. $Z_{bN}$ is the inversion of nodal admittance matrix $Y_{bN}$ \cite{gu2021impedance} rather than a simple branch impedance. $Z_b$ is dynamic impedance formulated in the $s$-domain (not simply $R+jX$) so all EMT dynamics are preserved. The entire procedure for port-mapping and port-based modelling are summarized in the model derivation tree in \figref{Fig:ModelDerivationTree} for a clearer review.

\subsection{Port-Based System Bipartition} \label{Section:PortAnalysis}

Now we explain the benefits of the port-mapping approach in whole-system stability studies. System bipartition is commonly used to investigate the interaction of sub-systems in a power system. For example, in the electrical-centric view, a power system is separated into sources and loads and their interaction is characterised by the impedance ratio between the source part and load part. This bipartition method can be naturally generalized in the port-based approach which turns out to provide additional flexibility in analyzing dynamic interaction from different views (mechanical-centric and electrical-centric). 

\begin{figure}[t!]
\centering
\includegraphics[scale=0.9]{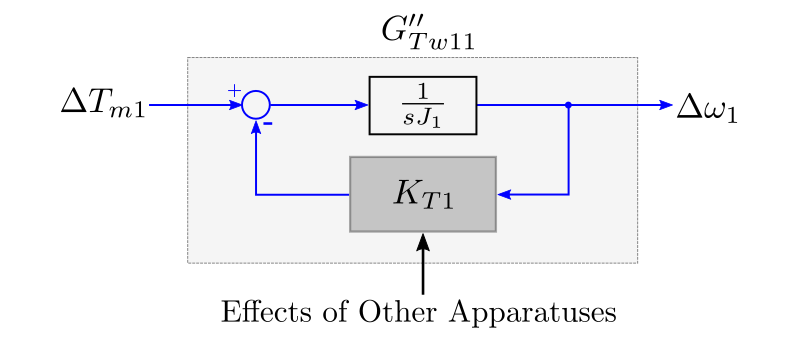}
\caption{The bipartite system viewed from mechanical port of SG1.}
\label{Fig:PortModel_Tw}
\end{figure}

For instance, we can split the system in \figref{Fig:PortModel_Tw} at the $T\text{-}\omega$ (mechanical) port of SG1. The transfer function seen at this port is ${G_{T\omega11}^{\prime\prime}}$. We take out the rotor inertia from this port and model the dynamics the rest parts of the grid as a dynamic torque coefficient $K_{T1}$
\begin{equation}
K_{T1} = {G_{T\omega11}^{\prime\prime\, -1}} - s J_1.
\end{equation}
The stability of the system is boiled down to the closed-loop interaction between rotor inertia $\frac{1}{sJ_1}$ and the dynamic torque coefficient $K_{T1}$, as shown in \figref{Fig:PortModel_Tw}. $K_{T1}$ is an extension to the classic torque coefficient and it encodes the dynamics of the entire grid (including EMT) except the rotor of SG1. If $K_{T1}$ itself is stable, the whole-system stability can be evaluated by the loop gain of \figref{Fig:PortModel_Tw} via Nyquist criterion. Since $\frac{1}{sJ_1}$ has a constant phase shift of $-90^\circ$ along the Nyquist contour, the phase of $K_{T1}$ should be kept above $-90^\circ$ to ensure positive phase margin, which leads to a very simple stability index.

\begin{figure}[t!]
\centering
\includegraphics[scale=0.9]{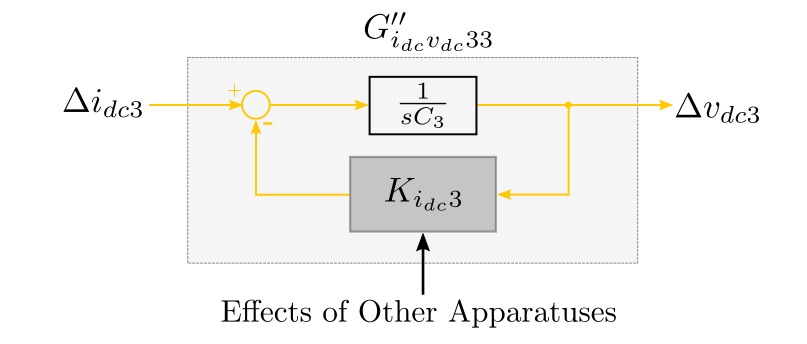}
\caption{The bipartite system viewed from dc-link port of IBR3.}
\label{Fig:PortModel_dc}
\end{figure}

Similarly, we can also split the system in \figref{Fig:PortModel_Tw} at the $i_{dc}\text{-}v_{dc}$ (dc-link) port of IBR3. We define the dc-link capacitance as the \textit{electrical inertia} and takes it out from the port transfer function $G_{i_{dc}v_{dc}33}^{\prime\prime}$ to yield the closed-loop diagram in \figref{Fig:PortModel_dc} where
\begin{equation}
    K_{i_{dc}3} = {G_{i_{dc}v_{dc}33}^{\prime\prime\,-1}} - s C_3
\end{equation}
is called the \textit{dc-current coefficient}.
$K_{i_{dc}3}$ plays a similar role to $K_{T1}$ which are named \textit{generalised torque coefficient} together.

It is possible to explore other ways of bipartition at different ports, e.g. an exciter port of a SG, or a $V\text{-}Q$ control port of an IBR. The port under investigation can be select to the point where oscillations stand out to provide insightful interpretations right on the spot. We can also split multiple ports simultaneously to investigate oscillations in which multiple apparatuses are participating.  

\subsection{Practical Implementation}

The port-based method proposed in this paper can be formulated in either state-space or transfer-function format, although we use transfer functions in our formulation, mainly for brevity in illustration. \revision{Inspire by the widely used open-source tools which automate system analysis work flow \cite{henriquez2020lits,cole2010matdyn,thurner2018pandapower}, we developed our own open-source toolbox to automate the proposed port-based stability analysis \cite{FuturePowerNetworks}. User only need to input a simple table specifying network layout and apparatus parameters, and the toolbox assemble the port-based whole-system model and visualise the stability results in time domain and frequency domain in just one click. The assembled models are graphically presented as Simulink block diagrams.}

\revision{As pointed out in \sectionref{Section:PortCoupling_Frame}, the proposed port-mapping method eliminates the global synchronisation signals (maps them into ports) and thus is a completely modular approach. Modular stabilization techniques, such as the passivity-based method \cite{harnefors2016passivity,gu2015passivity}, scale-free synthesis \cite{pates2019robust}, and port-Hamilton method \cite{zonetti2016energy}, can thus be easily incorporated in our port-based model. By contrast, prior-art component connection model \cite{wang2018small} cannot directly makes use of these techniques as the global synchronisation signals break modularity.}

% =========================
% Section: Case Study
% =========================

\section{Case Studies} \label{Section:CaseStudy}

In this section, the effectiveness of port-mapping modeling is illustrated through analyzing system stability in \highlight{three} cases. The first two case, (a) single-SG-infinite-bus system and (b) single-IBR-weak-grid system revisit well-known stability concerns in very simple set-ups but with new insights. The third case, (c) modified IEEE 14-bus SG-IBR composite system, illustrates modelling of composite systems. The parameters and models of all three cases, generated from the procedures in \sectionref{Section:PortCouplingMethod} and \sectionref{Section:GeneralPortCouplingMethod}, are available at \cite{FuturePowerNetworks}.

\subsection{Case 1: Single-SG-Infinite-Bus System} \label{Section:Case_SG}

\begin{figure}[t!]
\centering
\includegraphics[scale=0.7]{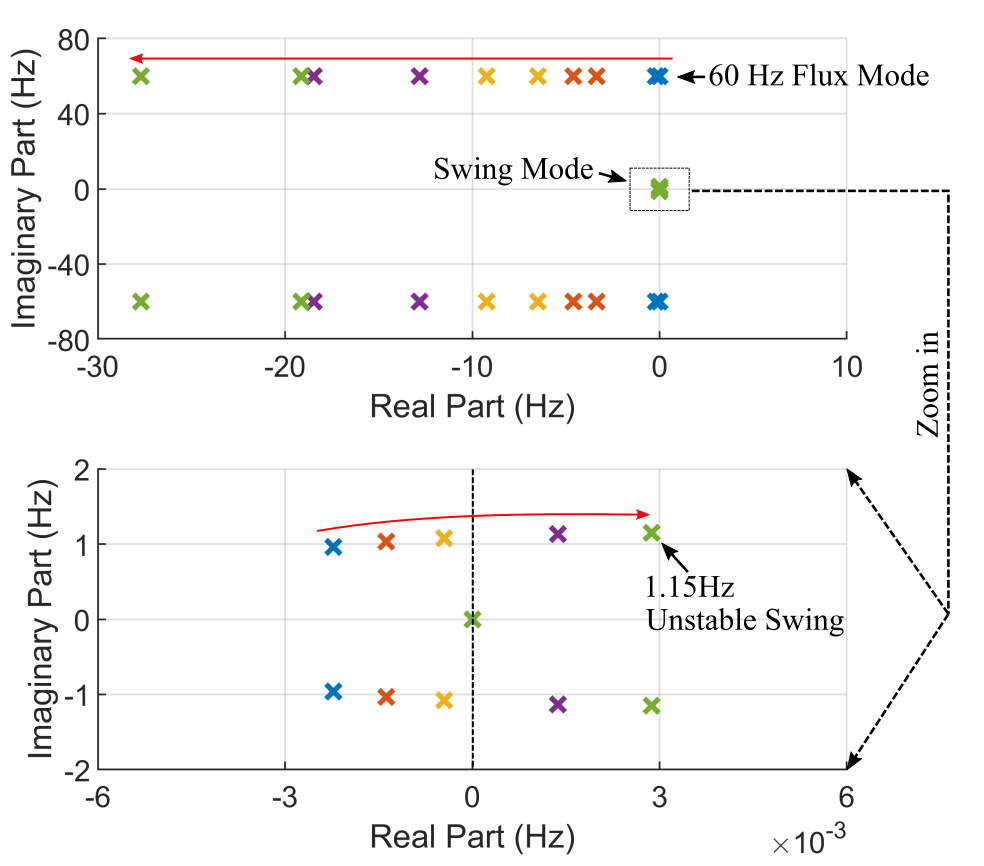}
\caption{Case 1: whole-system pole locus with increasing $R_b$ from \highlight{0 pu} to \highlight{0.3 pu}.}
\label{Fig:Case_SG_PoleLocus}
\end{figure}

\begin{figure}[t!]
\centering
\includegraphics[scale=0.8]{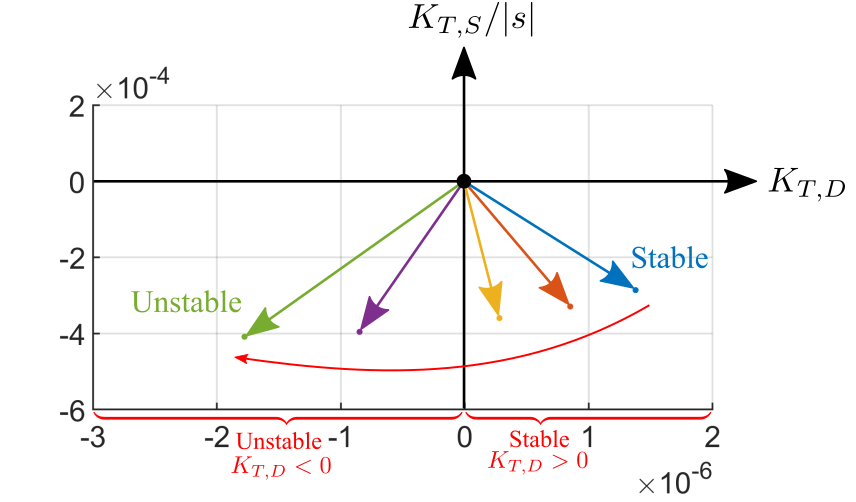}
\caption{Case 1:  vector diagram of torque coefficient $K_T$ when increasing the line resistance from \highlight{0 pu} to \highlight{0.3 pu}.}
\label{Fig:Case_SG_Complex_K}
\end{figure}

A single-SG-infinite-bus system and the familiar swing-dynamics are used to illustrate the effect of changing the damping at the stator electrical port, i.e., the line resistance $R_b$, on the dynamics at the rotor mechanical port. The poles of whole-system model $G_{N}^{\prime\prime}$ are shown in \figref{Fig:Case_SG_PoleLocus} including a sweep of $R_b$ from \highlight{0 pu} to \highlight{0.3 pu}. The \highlight{60 Hz} electrical flux modes are damped and shifted leftward when $R_b$ increases, as expected because $R_b$ is in series with the inductor of the relevant flux. The zoomed-in pole locus in the lower part of \figref{Fig:Case_SG_PoleLocus} shows that the swing modes, in contrast, move to rightward: the increase of electrical damping through $R_b$ decreases the damping of the swing oscillation, i.e., decreases the mechanical damping, eventually causing instability when the SG's own damping coefficient $K_D$ is inadequate (at \highlight{0.2 pu} here). This can be further examined using the port bipartition method in \figref{Fig:PortModel_Tw} by calculating the torque coefficient $K_T$ inclusive of network effects. As shown in \figref{Fig:Case_SG_Complex_K}, as $R_b$ is increased from \highlight{0 pu} to \highlight{0.3 pu}, the vector $K_T$ moves into the third quadrant with its phase decreasing to $<-90^{\degree}$ which implies negative $K_{T,D}$ and an unstable swing. \revision{This variation in the angle of $K_T$ with $R_b$ is an example of translating the grid impedance into the torque coefficient through the bidirectional mapping of dynamics between electrical and mechanical ports, as discussed in theoretical terms in \sectionref{Section:PortCoupling_SG_Grid}. Here, increasing the positive electrical damping (line resistance $R_b$) is seen to map to a decreased mechanical damping (decreased dynamic damping torque coefficient $K_{T,D}$). This also implies that increasing the self mechanical damping of the SG can re-stabilize this oscillation as tested next.}

\begin{figure}[t!]
\centering
\includegraphics[scale=0.75]{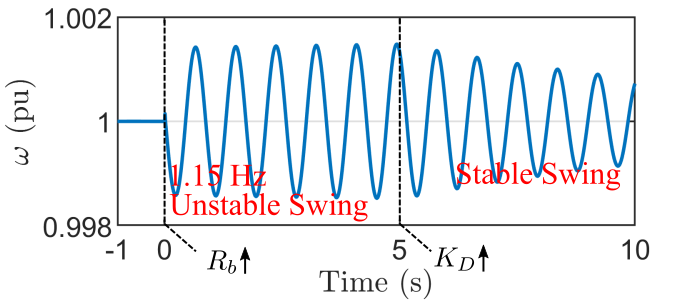}
\caption{Case 1: simulation results of SG dynamics. At \highlight{0 s}, line resistance $R_b$ is increased from \highlight{0 pu} to \highlight{0.3 pu} which destabilizes the swing mode. At \highlight{5 s}, $K_D$ of SG is increased from \highlight{0.2 pu} to \highlight{2 pu} which re-stabilizes the swing mode.}
\label{Fig:Case_SG_Simulation}
\end{figure}

The time-domain simulation results of \figref{Fig:Case_SG_Simulation} confirm that when line resistance $R_b$ is increased from \highlight{0 pu} to \highlight{0.3 pu} at $0~s$ leads to a \highlight{1.15 Hz} swing oscillation which is damped when at \highlight{5 s}, the mechanical damping $K_D$ of the SG is increased from \highlight{0.2 pu} to \highlight{2 pu}.

\subsection{Case 2: Single-IBR-Weak-Grid System} \label{Section:Case_IBR}

\begin{figure}[t!]
\centering
\includegraphics[scale=0.7]{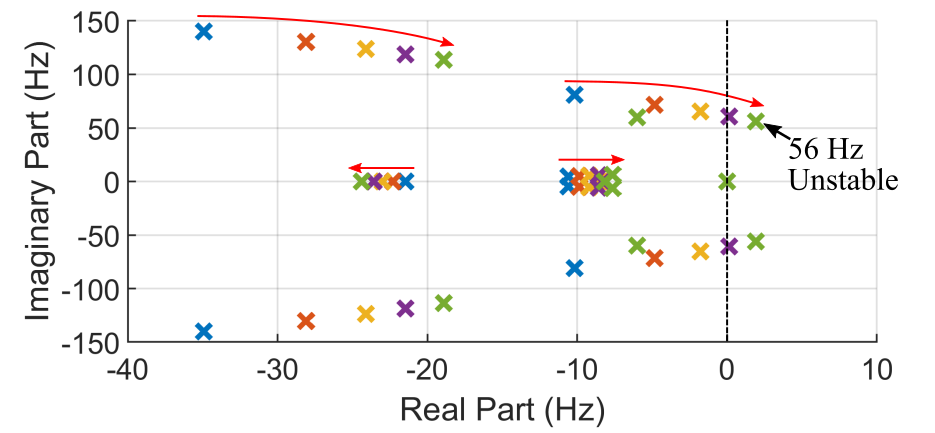}
\caption{Case 2: whole-system pole locus when reducing the SCR from \smallrevision{2.71} to \smallrevision{1.35}.}
\label{Fig:Case_IBR_PoleLocus}
\end{figure}

\begin{figure}[t!]
\centering
\includegraphics[scale=0.8]{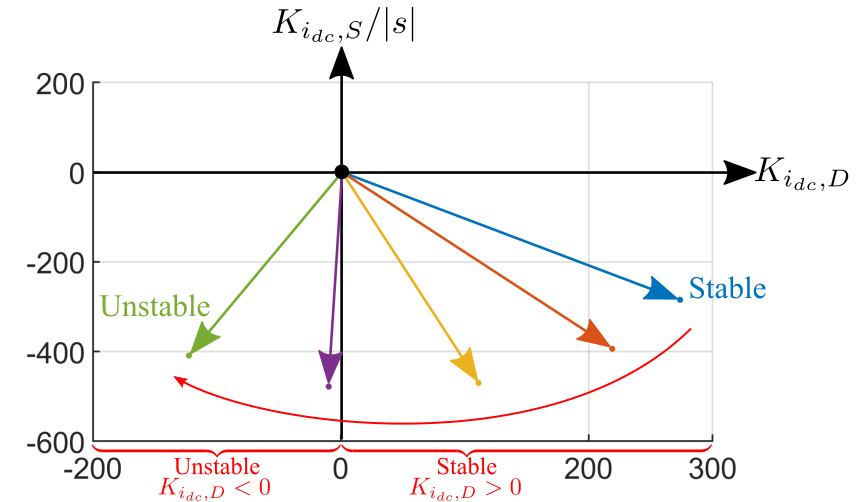}
\caption{Case 2: vector diagram of dc-current coefficient $K_{i_{dc}}$ when reducing the SCR from \smallrevision{2.71} to \smallrevision{1.35}.}
\label{Fig:Case_IBR_Complex_K}
\end{figure}

It is known that the stability of a grid-following IBR is degraded when connected to a weak grid, i.e., a grid with large impedance and low short-circuit ratio (SCR) \smallrevision{\cite{huang2015modeling}}. The port-separation method will again be used to identify how changed at the ac electrical port mapped to the dc-link port. The whole-system pole map in \figref{Fig:Case_IBR_PoleLocus} illustrates that when the SCR was reduced from \smallrevision{2.71} to \smallrevision{1.35}, the \highlight{56 Hz} mode became unstable. The plot of the dc-current coefficient vector $K_{i_{dc}}$ in \figref{Fig:Case_IBR_Complex_K} reveals the same onset of instability as a change in the angle of the vector. \revision{In this case, the changed dynamics caused by the high line impedance (low SCR) at the ac electrical port are seen to be mapped to the dc-current coefficient at dc electrical port of the IBR. This confirms that the port-mapping and port-separation methods can represent and reveal causes of IBR instability.}

\begin{figure}[t!]
\centering
\includegraphics[scale=0.75]{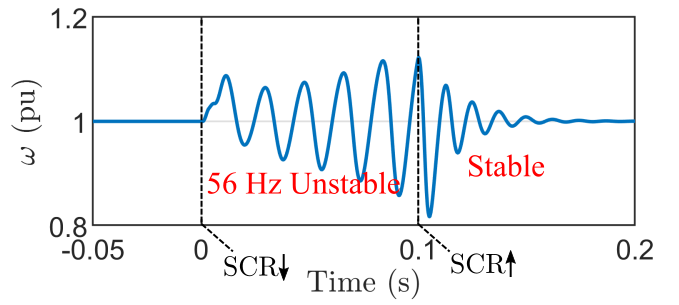}
\caption{Case 2: simulation results of IBR dynamics. At \highlight{0 s}, SCR is reduced from \smallrevision{2.71} to \smallrevision{1.35} which leads to the instability. At \highlight{0.1 s}, SCR is increased back to \smallrevision{2.71} which re-stabilizes the system.}
\label{Fig:Case_IBR_Simulation}
\end{figure}

The time-domain simulation results of \figref{Fig:Case_IBR_Simulation} confirm that when the gird is weakened with SCR reduced from \smallrevision{2.71} to \smallrevision{1.35} at \highlight{0 s}, a \highlight{56 Hz} unstable oscillation begins. At \highlight{0.1 s}, SCR is increased back to \smallrevision{2.71}, which re-stabilizes the system.

\subsection{Case 3: Modified IEEE 14-Bus SG-IBR Composite System} \label{Section:Case_Composite}

\begin{figure}[t!]
\centering
\includegraphics[scale=0.6]{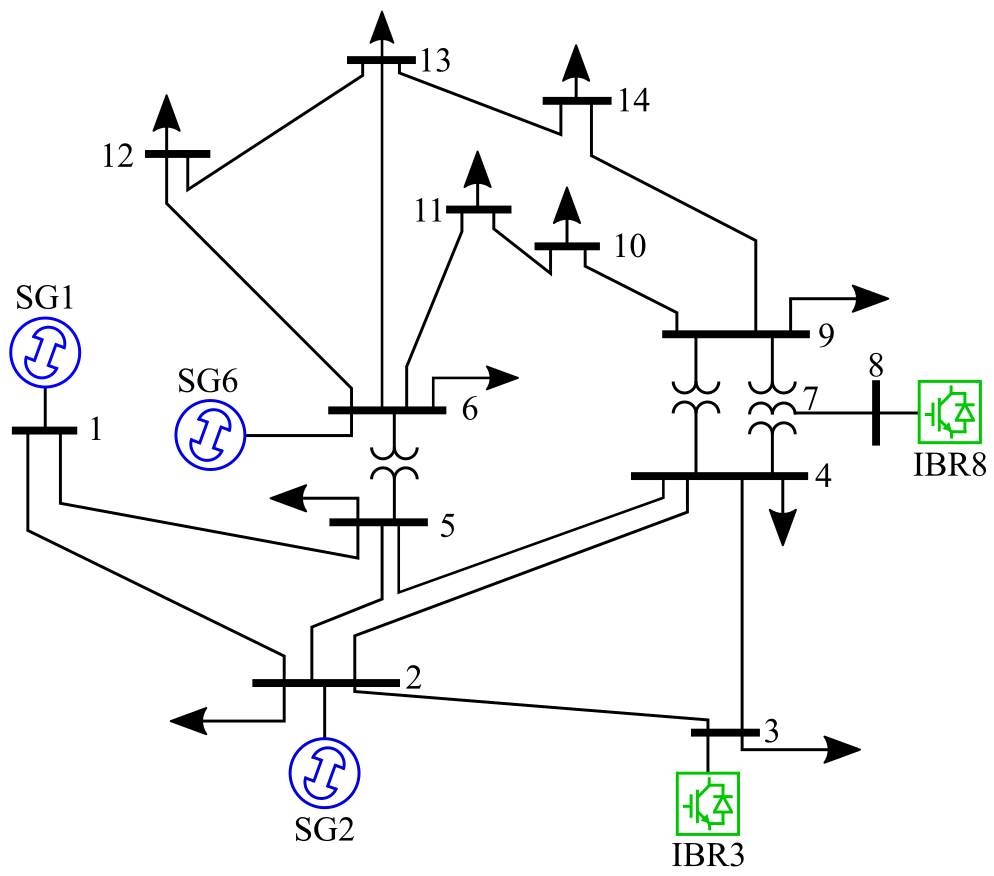}
\caption{Case 3: IEEE 14-bus SG-IBR composite power system.}
\label{Fig:Case_Composite_14Bus}
\end{figure}

\begin{figure}[t!]
\centering
\includegraphics[scale=0.7]{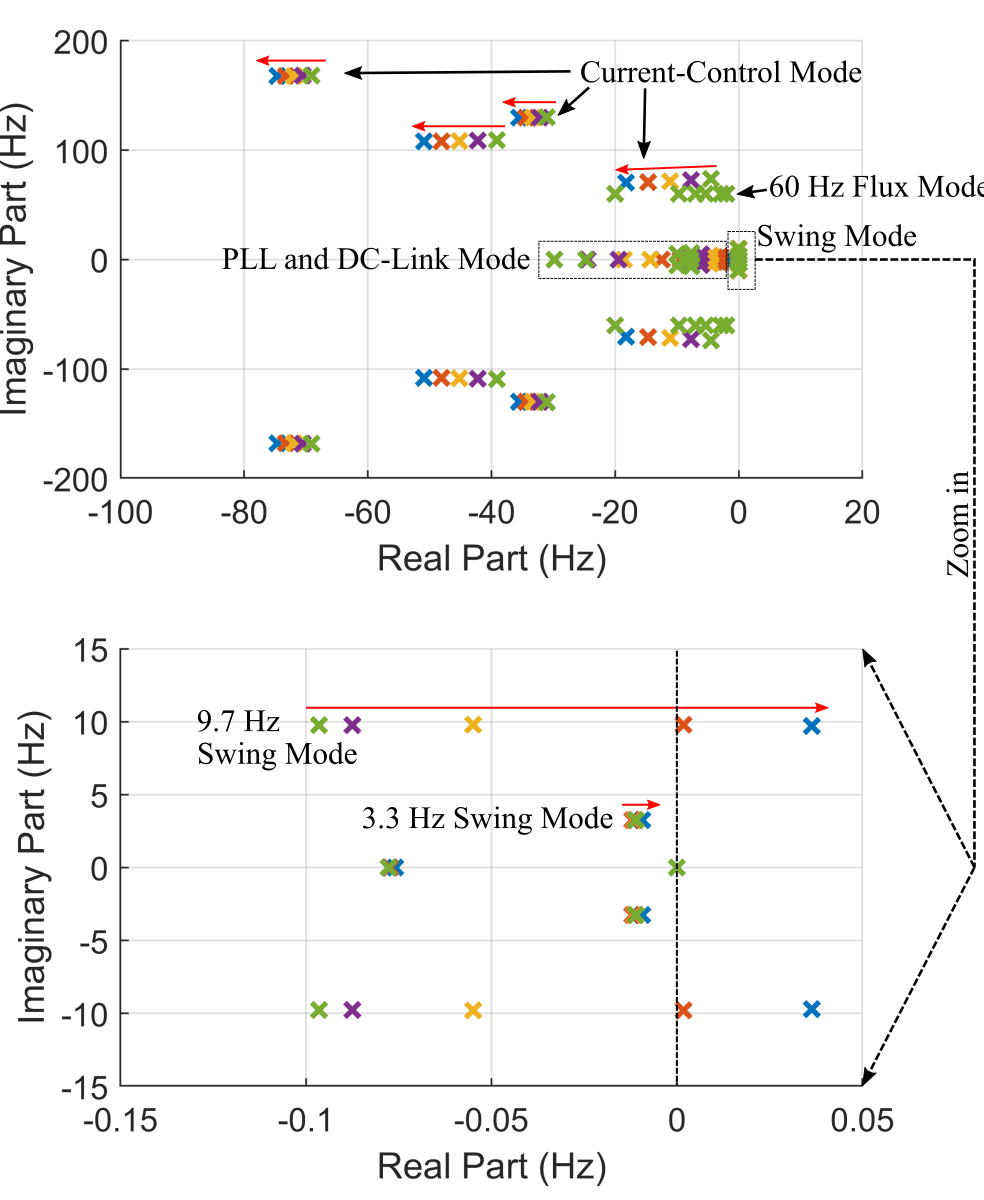}
\caption{Case 3: whole-system pole locus with reducing the IBR power-frequency-related control bandwidth (dc-link and PLL) from \highlight{25 Hz} to \highlight{5 Hz}.}
\label{Fig:Case_Composite_PoleLocus}
\end{figure}

\begin{figure}[t!]
\centering
\includegraphics[scale=0.7]{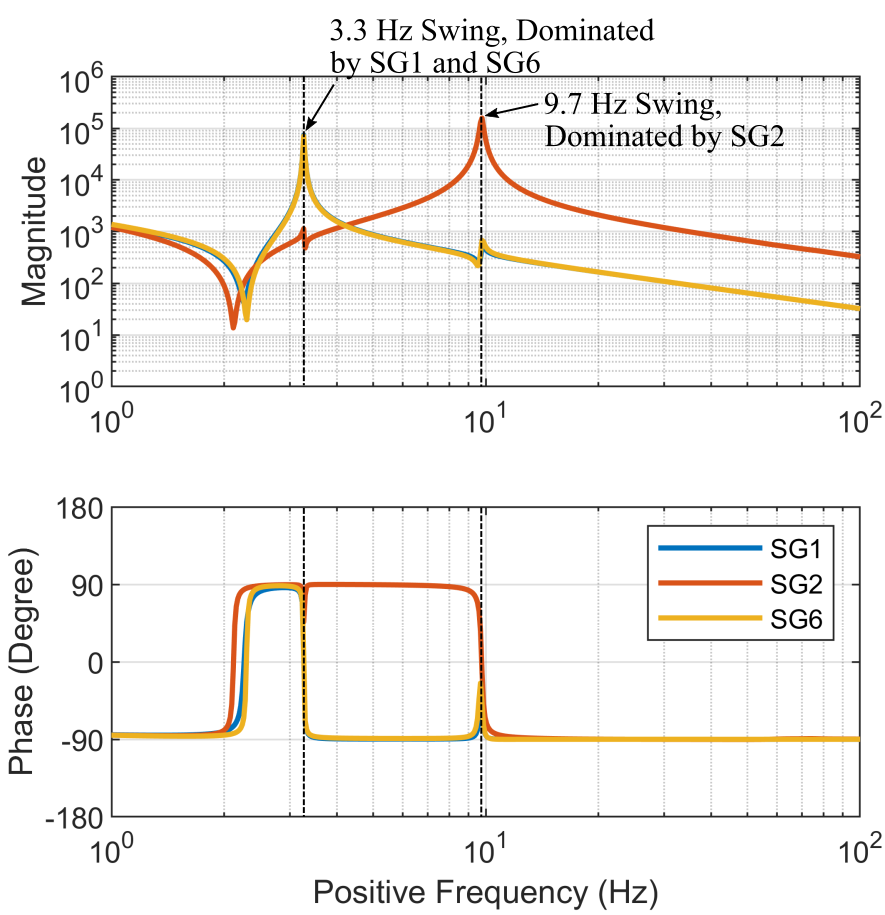}
\caption{Case 3: Bode diagram of $T\omega$ port closed-loop transfer function (i.e., the mechanical impedance $G_{T\omega}^{\prime\prime}$ in $G_{N}^{\prime\prime}$) for each SG.}
\label{Fig:Case_Composite_Bode_Tw}
\end{figure}

\begin{figure}[t!]
\centering
\subfloat[]{\includegraphics[scale = 0.7]{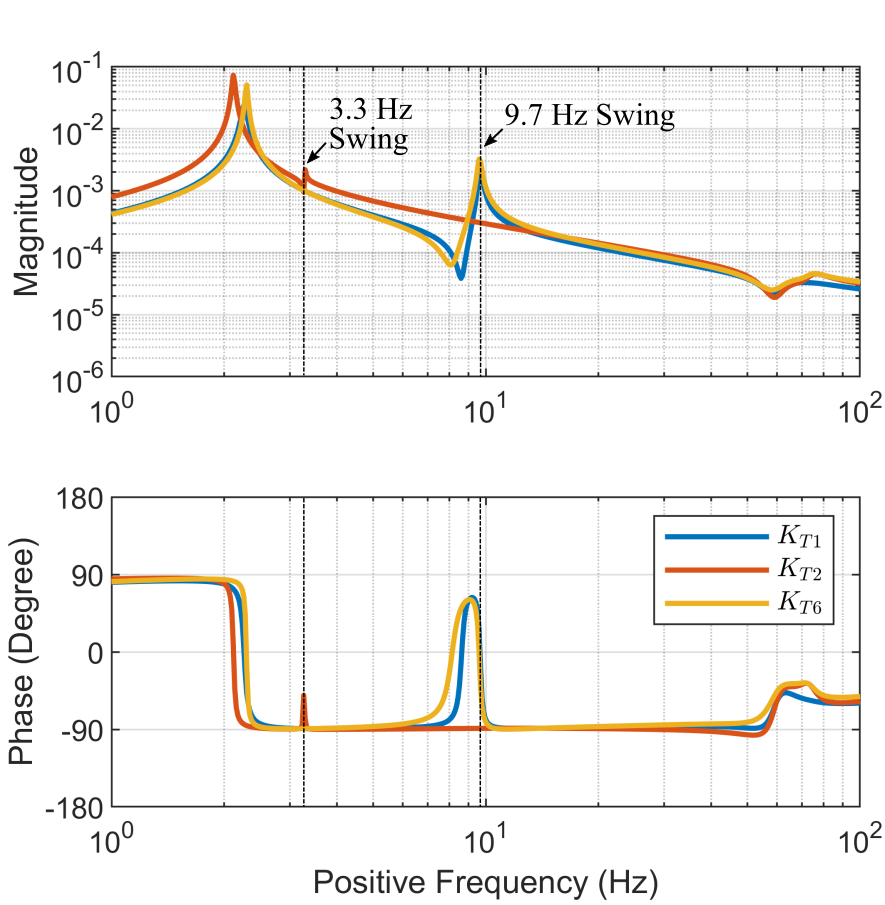}}

\subfloat[]{\includegraphics[scale = 0.7]{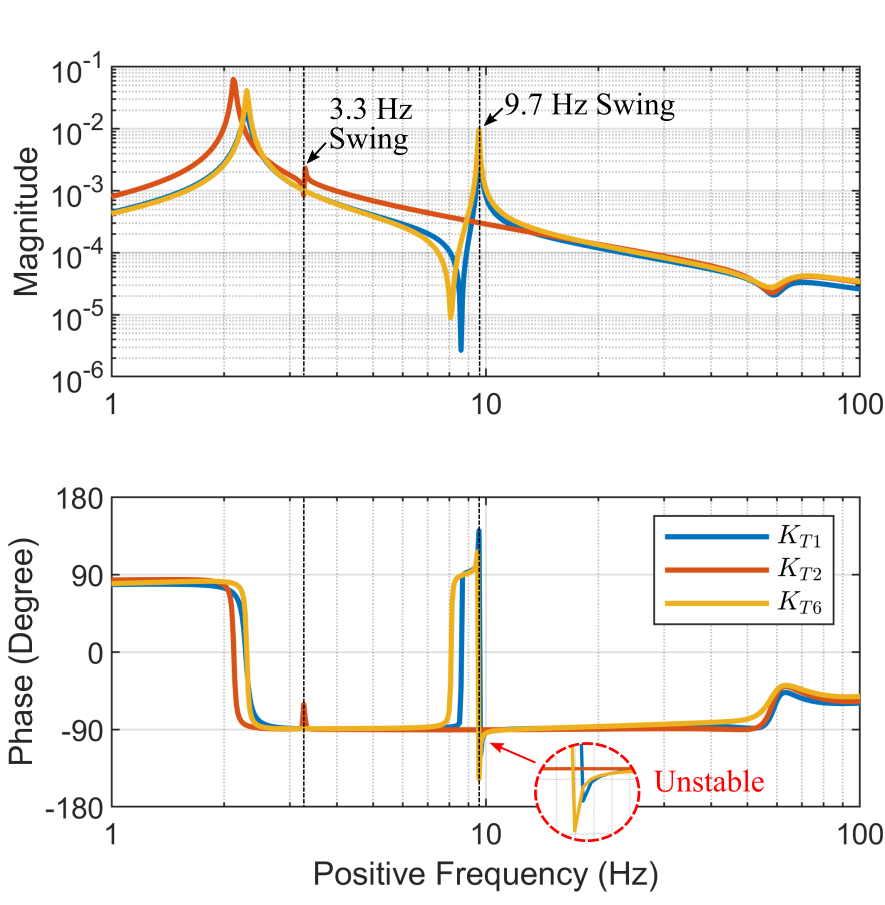}}
\caption{Case 3: Bode diagram of torque coefficient $K_T$ for each SG with changing the power-frequency control bandwidth (dc-link and PLL) of IBRs. (a) Bandwidth of \highlight{25 Hz}. (b) Bandwidth of \highlight{5 Hz}.}
\label{Fig:Case_Composite_Bode_K}
\end{figure}

To analyze the power-frequency interaction between SGs and IBRs, a modified IEEE 14-bus power system is investigated next, as shown in \figref{Fig:Case_Composite_14Bus}. The same layout of the IEEE standard 14-bus system is retained but two SGs are replaced by IBRs. \figref{Fig:Case_Composite_PoleLocus} shows the poles of the whole system when bandwidth of the power-frequency controllers (i.e., dc-link and PLL) of the IBRs are reduced from \highlight{25 Hz} to \highlight{5 Hz}. The zoomed-in figure shows the loci of the swing modes. These can be classified into two groups: stable \highlight{3.3 Hz} poles and unstable \highlight{9.7 Hz} poles. \revision{The magnitude of resonant peak of closed-loop gain $G_{T\omega}^{\prime\prime}$ for each SG in \figref{Fig:Case_Composite_Bode_Tw} reveals the participation of apparatuses on oscillation modes \cite{zhu2021participation}. It indicates that the \highlight{9.7 Hz} swing mode is dominated by SG2 (red line), and the \highlight{3.3 Hz} swing mode is jointly dominated by SG1 (blue line) and SG6 (yellow line).} The port-separation method can be used here to obtain the torque coefficient $K_N$ of each SG, as shown in \figref{Fig:Case_Composite_Bode_K}, where the bode plot is used here so that we can investigate the magnitude and phase of the torque coefficient vectors at both swing mode frequencies in same figure. Results for control bandwidths of \highlight{25 Hz} are in \highlight{(a)} and \highlight{5 Hz} in \highlight{(b)}. A phase of $K_T$ smaller than $-90^{\degree}$ can be observed at the \highlight{9.7 Hz} swing point for bandwidth of \highlight{5Hz} which agrees with the pole loci.

\begin{figure}[t!]
\centering
\includegraphics[scale=0.75]{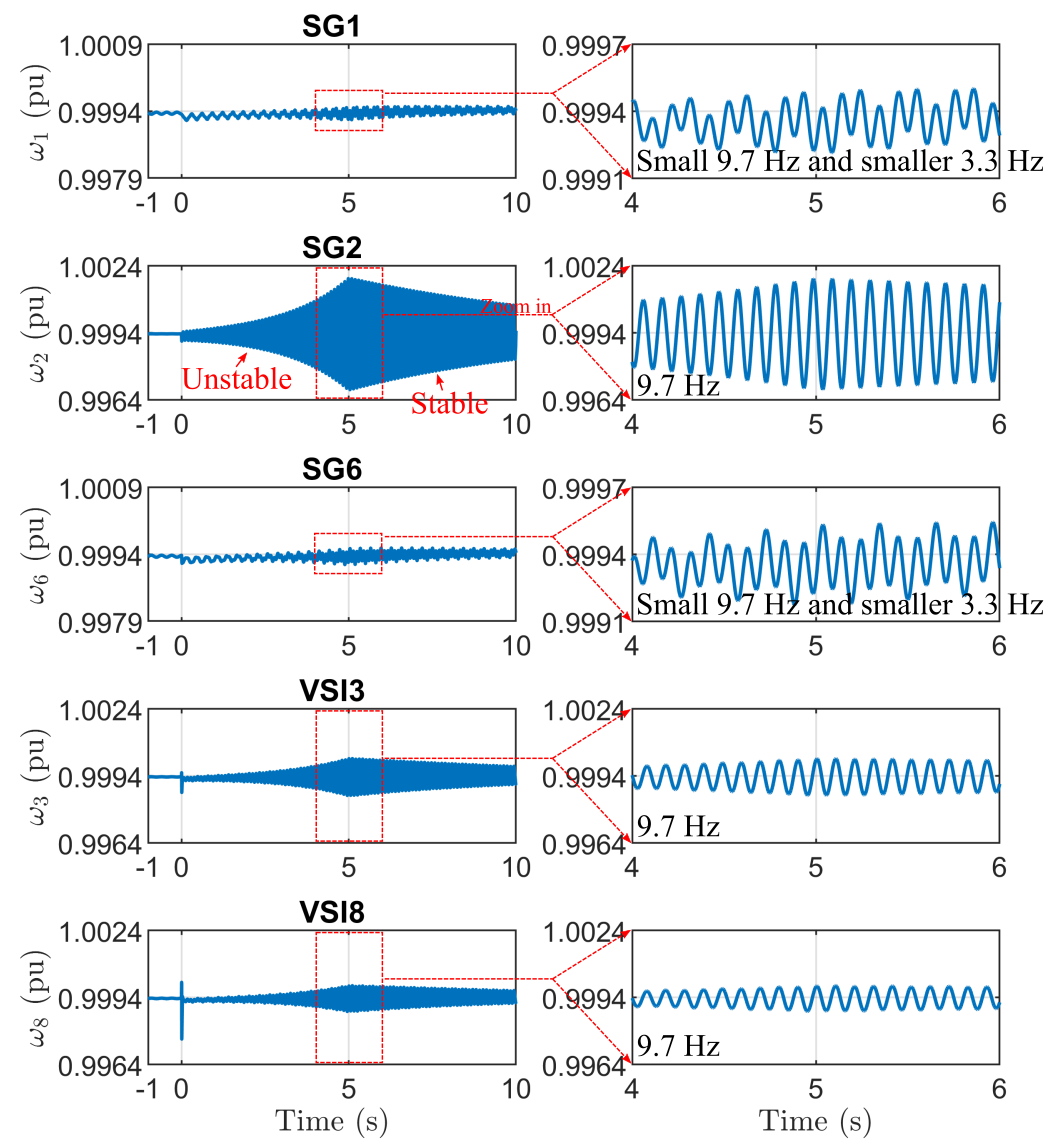}
\caption{Case 3: simulation results. At \highlight{0 s}, the power-frequency control bandwidth of IBR (dc-link and PLL) is reduced from \highlight{25 Hz} to \highlight{5 Hz}, which destabilizes the \highlight{9.7 Hz} swing mode. At \highlight{5 s}, the damping torque coefficient $K_{D2}$ of SG2 is increased from \highlight{1 pu} to \highlight{2 pu}, which re-stabilizes the system.}
\label{Fig:Case_Composite_Simulation}
\end{figure}

The corresponding time-domain simulation results are shown in \figref{Fig:Case_Composite_Simulation}. At \highlight{0 s}, the IBR power-frequency bandwidth is reduced from \highlight{25 Hz} to \highlight{5 Hz} and the unstable \highlight{9.7 Hz} swing oscillation begins. 
%The severe oscillations of SG2, IBR3, and IBR8 imply that these are the sources of the instability rather than SG1 and SG3, which is as predicted by the theoretical analysis. 
At \highlight{5 s}, the damping torque coefficient $K_{D2}$ of the critical SG2 is increased from \highlight{1 pu} to \highlight{2 pu}, which re-stabilizes the system.

% =========================
% Section: Conclusions
% =========================

\section{Conclusions} \label{Section:Conclusions}

A port-mapping method has been put forward for stability analysis of composite power systems with mixtures of SGs and IBRs. The proposed method maps the dynamics between multi-physical-domain ports and gives more insights on whole system stability. It unifies the mechanical- and electrical-centric analysis of power systems in a single model. The mechanical rotational dynamics are mapped to the electrical impedance or admittance, and meanwhile the electrical dynamics are also mapped to the mechanical torque coefficient. This united cross-domain methodology enables more flexible bipartition of power systems to allow for more interpretative investigation into the interaction amongst sub-systems, which yields new stability indices such as generalised torque coefficients.

% =========================
% Section: Appendix
% =========================

\appendices

\section{State Equations of Synchronous Generator} \label{Appendix:Equ_SG}

State equations of a SG with considering both stator electrical dynamics and rotor mechanical dynamics (in \smallrevision{source} convention) are \cite{kundur1994power,gu2021impedance}
\begin{equation} \label{Equ:SG}
\begin{aligned}
v_{d} & = \dot{\psi}_{d} - R i_{d} - \omega \psi_{q}\\
v_{q} & = \dot{\psi}_{q} - R i_{q} + \omega \psi_{d}\\
J\dot{\omega} & = T_{m}- T_{e} - K_D{\omega}\\
\end{aligned}
\end{equation}
with
\begin{equation} \label{Equ:SG_T}
\psi_d = - L i_d,~\psi_q = - L i_q - \psi_f,~T_e = \psi_f i_d
\end{equation}
where $\psi_d$ and $\psi_q$ are $dq$-axis flux linkages respectively; $\psi_f$ is the field flux linkage; $v_d$ and $v_q$ are $dq$-axis stator voltages; $i_d$ and $i_q$ are $dq$-axis stator currents; $\omega$ is the rotor speed; $J$ is the rotor inertia; $T_m$ and $T_e$ are the mechanical and electromagnetic torque respectively; $K_D$ is the damping torque coefficient; $R$ and $L$ are stator resistance and inductance respectively.

\section{Model Derivation of Port-Mapping Method} \label{Appendix:Step}

\subsection{$G$ (Mechanical Frame) $\rightarrow$ $G^\prime$ (Global Frame)} \label{Appendix:Step_Frame}

According to \figref{Fig:SingleMachineInfiniteBus} (e), the element relationship between $G$ (mechanical frame) and $G^\prime$ (electrical frame) can be obtained as
\begin{equation} \label{Equ:FrameDynamicsEmbedding}
\begin{aligned}
& G_{vi}^{\prime} 
= 
T_{\epsilon_0}
\bigg(
G_{vi} + (I_0 - G_{vi} V_0)\underbrace{(s+G_{v\omega} V_0)^{-1}G_{v\omega}}_{G_{v\omega}^{\prime}T_{\epsilon_0}/s}
\bigg)
T_{\epsilon_0}^{-1}
\\ 
& G_{Ti}^{\prime}
= 
T_{\epsilon_0}
\bigg(
G_{Ti} + (I_0 - G_{vi} V_0)\underbrace{(s+G_{v\omega} V_0)^{-1}G_{T\omega}}_{G_{T\omega}^{\prime}/s}
\bigg)
\\
& G_{v\omega}^{\prime} 
= 
\bigg(
s(s+G_{v\omega} V_0)^{-1}G_{v\omega}
\bigg)
T_{\epsilon_0}^{-1}
\\ 
& G_{T\omega}^{\prime} 
= 
s(s+G_{v\omega} V_0)^{-1}G_{T\omega}
\end{aligned}
\end{equation}
where $I_0 = [-i_{q0};i_{d0}]$ and $V_0 = [-v_{q0};v_{d0}]$ are the steady-state operating points at the electrical port; $T_{\epsilon_0}$ is the frame transformation matrix with the steady-state angle $\epsilon_0$. The first equation in \equref{Equ:FrameDynamicsEmbedding} builds the relationship between $G_{vi}$ (admittance in mechanical frame) and  $G_{vi}^\prime$ (admittance in electrical frame), which correspond to the impedance-based analysis in \cite{gu2021impedance,li2020interpreting} albeit that they focus on the electrical domain only.
%If setting $G_{v\omega} = sK_v$ where $K_v$ is the gain from $\Delta v_{dq}$ to $\Delta \epsilon$, the relationship between $G_{vi}^\prime$ and $G_{vi}$ can be re-illustrated as
%\begin{equation}
%\begin{aligned}
%Y_{d^\prime q^\prime} 
%& = Y_{dq} + (I_0 - Y_{dq}V_0)(I + K_vV_0)^{-1}K_v 
%\\
%& = (Y_{dq}+I_0 K_v)(I+V_0 K_v)^{-1}
%\end{aligned}
%\end{equation}
% It coincides the impedance-based analysis in \cite{gu2021impedance}, which can be regarded as an inference of the derived model in this paper by focusing on the electrical domain only.

\subsection{$G^{\prime}$ (Apparatus Model) $\rightarrow$ $G^{\prime\prime}$ (Whole System Model)} \label{Appendix:Step_SG_Grid}

According to \figref{Fig:SingleMachineInfiniteBus} (e), the element relationship between $G^\prime$ (apparatus model) and $G^{\prime\prime}$ (whole system model for the single-SG-infinite-bus system) can be obtained as
\begin{equation} \label{Equ:G_Relationship_Impedance}
\begin{aligned}
& G_{vi}^{\prime\prime} = (I-G_{vi}^{\prime} Z_b)^{-1} G_{vi}^{\prime}
\\
& G_{Ti}^{\prime\prime} = (I-G_{vi}^{\prime} Z_b)^{-1} G_{Ti}^{\prime}
\\
& G_{v\omega}^{\prime\prime} = G_{v\omega}^{\prime} + G_{v\omega}^{\prime} Z_b \underbrace{(I - G_{vi}^{\prime} Z_b)^{-1} G_{vi}^{\prime}}_{G_{vi}^{\prime\prime}}
\\
& G_{T\omega}^{\prime\prime} = G_{T\omega}^{\prime} + G_{v\omega}^{\prime} Z_b \underbrace{(I - G_{vi}^{\prime} Z_b)^{-1} G_{Ti}^{\prime}}_{G_{Ti}^{\prime\prime}}
\end{aligned}
\end{equation}

% =========================================================
% End
% =========================================================

\ifCLASSOPTIONcaptionsoff
  \newpage
\fi

\bibliographystyle{IEEEtran}
\bibliography{Paper_Port}

\end{document}